\documentclass[runningheads]{llncs}
\usepackage[T1]{fontenc}
\usepackage[pdftex, svgnames, dvipsnames]{xcolor}
\usepackage{graphicx}
\usepackage{comment}
\usepackage{caption}
\usepackage{subcaption}
\usepackage{float}
\usepackage{booktabs}
\usepackage{makecell}
\usepackage{listings}
\usepackage{pgfplots}
\usepackage{todonotes}
\usepackage{xspace}
\usepackage{url}
\usepackage{multirow}
\usepackage{subfig}
\usepackage{fixme}
\usepackage{adjustbox}
\usepackage[shortlabels]{enumitem}

\newcommand{\ie}{i.e., \@\xspace} 
\newcommand{\eg}{e.g., \@\xspace} 

\fxsetup{status=draft, layout=inline, theme=color}
\FXRegisterAuthor{ss}{ssa}{\colorbox{Purple}{\color{White}SS}}
\FXRegisterAuthor{ds}{dso}{\colorbox{Brown}{\color{White}DS}}
\FXRegisterAuthor{dm}{dma}{\colorbox{Violet}{\color{White}DM}}
\FXRegisterAuthor{gg}{ggi}{\colorbox{Periwinkle}{\color{White}GG}}
\FXRegisterAuthor{all}{alln}{\colorbox{Green}{\color{White}All}}

\begin{document}
\title{Are Trees Really Green? A Detection Approach of IoT Malware Attacks}
\titlerunning{Green Machine Learning for IoT Detection}
%

\author{Silvia Lucia Sanna\inst{1}\orcidID{0009-0002-8269-9777} \and
Diego Soi\inst{1}\orcidID{0009-0009-0092-9067} \and
Davide Maiorca\inst{1}\orcidID{0000-0003-2640-4663}\and
Giorgio Giacinto\inst{1,2}\orcidID{0000-0002-5759-3017}
}

\authorrunning{Sanna et al.}

\institute{Dip. Ingegneria Elettrica ed Elettronica, \\ Università degli Studi di Cagliari, Cagliari, Italy\and
CINI, Consorzio Interuniversitario Nazionale per l’Informatica, Roma, Italy\\
\email{\{silvial.sanna, diego.soi, davide.maiorca, giogio.giacinto\}}@unica.it
}



\maketitle              
\begin{abstract}

Nowadays, the Internet of Things (IoT) is widely employed, and its usage is growing exponentially because it facilitates remote monitoring, predictive maintenance, and data-driven decision making, especially in the healthcare and industrial sectors. However, IoT devices remain vulnerable due to their resource constraints and difficulty in applying security patches. Consequently, various cybersecurity attacks are reported daily, such as Denial of Service, particularly in IoT-driven solutions.

Most attack detection methodologies are based on Machine Learning (ML) techniques, which can detect attack patterns. However, the focus is more on identification rather than considering the impact of ML algorithms on computational resources.

This paper proposes a \emph{green} methodology to identify IoT malware networking attacks based on flow privacy-preserving statistical features. 
In particular, the hyperparameters of three tree-based models -- Decision Trees, Random Forest and Extra-Trees -- are optimized based on energy consumption and test-time performance in terms of Matthew’s Correlation Coefficient.

Our results show that models maintain high performance and detection accuracy while consistently reducing power usage in terms of watt-hours (Wh). This suggests that on-premise ML-based Intrusion Detection Systems are suitable for IoT and other resource-constrained devices.

\end{abstract}

\section{Introduction}\label{sec:intro}

Nowadays, Internet of Things (IoT) devices and their interconnections are becoming exponentially important in everyday life, from industry~\cite{Xu2014_TIF} to houses, vehicles, and smart cities of interconnected systems~\cite{Keertikumar2015_ICGIoT}. Such devices have low capabilities in terms of energy and computational resources compared to desktop and server computers. They are often employed to measure specific data, \ie \emph{sensors}, and control mechanical systems or forces, \ie \emph{actuators}. With the advent of Artificial Intelligence (AI), these devices are becoming more intelligent and capable of making decisions independently~\cite{Atzori2010_CN}.

Over the years, IoT devices and interconnected networks have been found to be susceptible to various vulnerabilities~\cite{Meneghello2019_IoTJ}, which have allowed several types of cybersecurity attacks, such as data exfiltration, Denial of Service (DoS), and its Distributed variant (DDoS). Most of the time, these attacks are perpetrated by malware specifically designed for IoT systems, which are generally equipped with limited operating systems (OS) based on the Linux kernel, such as {\tt Miraii}~\cite{Mirai}, and {\tt Torii}~\cite{Torii},  and {\tt Hide\&Seek}~\cite{HideSeek}. Detecting the presence of malware in an IoT environment is crucial as a first step to counteract a large number of cyberattacks, given the growing importance of IoT devices, their security disadvantages~\cite{Schiller2022_CSR}, and the limited resources they are provided with.

In this work, we focused on three detection strategies based on network communications made by IoT devices, which normally communicate with remote servers for data exchange. When malware takes control over the expected operability, the packet transmissions change, and attack patterns can be identified to detect anomalous behavior. These systems, called Intrusion Detection Systems (IDS), are generally deployed on-premise in ad-hoc devices, \ie firewalls or routers, which work as access points for the IoT devices, or they are host-based, running directly on the device under analysis. However, these systems are typically resource-consuming, especially if they rely on Machine Learning (ML) algorithms~\cite{Martin2019_PKDD}, which require significant computational power and memory, making them unsuitable for deployment on resource-constrained devices.

To the best of our knowledge, few works at the state of the art focus on \emph{Green Machine Learning} for network security applications~\cite{Ioannou2024_CC,Tekin2023_IoT}, that is, optimizing the underlying ML model to find a balance between drained energy and performance. Indeed, a green ML strategy is necessary mainly for two aspects. First, maintaining high detection performance while reducing consumed energy allows the deployment of IDS on-premise directly in IoT devices. Second, the longer an attack remains undetected, the greater the energy consumption will be due to malware operation and legitimate activities the IoT device may perform. As a consequence, the operability lifecycle of the device diminishes, causing the so-called \emph{e-waste}~\cite{Modarress22_Sust} of electronics. Therefore, reducing the size of detection algorithms and the consumed energy would increase the device's life. Additionally, reducing the overall energy consumed is fundamental to decreasing the equivalent carbon footprint, helping to combat climate change.

In this paper, we employed a dataset of common network cyberattacks in different IoT scenarios~\cite{IoT23-dataset} (\eg DoS and port-scanning) to identify post-mortem anomalies based on network features. We optimized the hyperparameters of three tree-based ML algorithms to minimize energy consumption during the testing phase while maximizing performance. We demonstrate that accuracy is not significantly affected by lower power usage. Our approach can be seen as an adaptive, energy-efficient IDS designed for network traffic detection.

To this end, we sketch two modes of operation: \begin{enumerate}[a)] 

\item \emph{non-green}, \ie the training phase, in which the ML algorithm is optimized on a server, enabling fast learning and handling large training datasets, while still prioritizing power usage reduction during testing; 
\item \emph{green} mode during runtime in resource-constrained devices, triggering alerts when an anomaly is detected as in the testing phase. 
\end{enumerate}

In this way, the algorithm is trained with large datasets to account for most learning patterns, while the algorithm running on-premise lowers power consumption, maintaining similar accuracy.

In summary, \emph{i}) we developed a detection methodology based on networking post-mortem features, optimizing the ML algorithms for energy consumption in terms of $\mu$Wh and performance during the testing phase; \emph{ii}) we selected only statistical features per flow without considering the body of the packets, \ie a \emph{privacy-preserving} approach; \emph{iii}) we compared different ML algorithms to understand which is the most \emph{green} during the testing phase; and \emph{iv}) we defined the importance of false negative flows not detected by the system.

The rest of the paper is organized as follows. Section~\ref{sec:sota} reviews the literature on detecting malware networking communication and the energy consumption of ML systems, and Section~\ref{sec:methodology} describes the dataset, features, and employed algorithms. Section~\ref{sec:results} discusses the results of the classification with respect to the performance and energy consumption of the models. Finally, Section~\ref{sec:conclusion} discusses the limitations of the approach and future works.

\section{Related Works}\label{sec:sota}


This section reviews recent works on current advancements in energy estimation with respect to AI technologies in Section~\ref{sec:sota:energy_measurement}, and the state-of-the-art regarding the detection of IoT malware through traffic patterns in Section~\ref{sec:sota:c2cdetection}.

\subsection{Energy Consumption Measurement}\label{sec:sota:energy_measurement}

Advancements in computer science technology powered by Cloud Computing and AI have substantially increased the demand for energy resources. This rise in energy consumption affects the feasibility of implementing cybersecurity solutions in battery-powered systems like IoT and mobile environments, while also contributing to $CO_2$ emissions, exacerbating environmental challenges such as climate change. Thus, the reduction of software energy consumption is becoming an interesting topic for the research community~\cite{Candelieri2024_ML,Yokoyama22_AAI,Yarally23_CAIN}. A recent survey~\cite{GarciaMartin2019_JPDC} reviewed two techniques for estimating the energy consumption of algorithms to better design software: \emph{i}) \emph{hardware-level} to compute the energy efficiency of hardware components (\ie CPU, RAM, and I/O peripherals); \emph{ii}) \emph{software-level} through simulation or real-time estimation at the instruction level to trace the consumed energy by performance counter profiling or instruction-set simulation.

Among the most recent approaches, Budennyy et al. proposed {\tt Eco2AI}~\cite{Budennyy2022_DM}, a framework that measures energy consumption in terms of Joules or KWh, focusing on CPU, GPU, and RAM real-time evaluation. {\tt PyJoule}\footnote{\url{https://github.com/powerapi-ng/pyJoules}} employs the {\tt Intel RAPL} (Running Average Power Limit) technology to estimate the power consumption of CPU, RAM and integrated GPU. Additionally, Antony et al. proposed {\tt Carbon Tracker}, a tool to track the energy and carbon footprint of ML models. The authors evaluated the tool's efficiency by comparing the estimations with the actual measurements done in monitoring $1$ training epoch for two models, \ie CNN and Autoencoders, with errors between $5\%$ and $19\%$. Due to its efficiency and attested good results, in this work we based the optimization of detection algorithms on the measurements done by {\tt Carbon Tracker} as explained in Section~\ref{sec:methodology}.

Other works, besides measuring the energy efficiency of neural network models and their training, suggest ways to reduce consumption. Tipp et al.~\cite{Tripp2024_arxiv} suggest, among other methods, reducing idle time when accessing memory to eliminate excess energy due to idle power drawn and reducing memory access by using specialized hardware to hold larger parameters in cache.

\subsection{Detection of IoT Malware Traffic}\label{sec:sota:c2cdetection}

The rapid evolution of IoT devices has attracted malware authors interested in exploiting security vulnerabilities through malicious software, whose aim is generally to gain unauthorized privileges in the network to which IoT devices are connected, like in the case of {\tt Mirai}~\cite{Mirai} and {\tt SILEX}~\cite{Mukthar2024_Symm} attacks. For this reason, one of the main approaches to identify this kind of malicious activity is to leverage the network patterns they generate.

One of the first works was published by Bilge et al.~\cite{Bilge12_ACSAC} in 2012, whose main goal was to propose invariant network features without considering the application protocols due to differences in each client/server communication. In particular, they selected flow size-based, client access pattern-based features, and temporal ones to characterize the variability of client flow volume as a function of time.
Recently, Davanian et al.~\cite{Davanian24_AsiaCCS} proposed a methodology to identify live C\&C servers with zero-priori knowledge to separate the C2-bound traffic from other traffic accurately. Their methodology is based on a SYN-DATA-aware approach, depending on the number of SYN flags and the data exchanged. Moreover, they focus on a grammar-based representation of the traffic, considered as a dialog, to create a fingerprint-aware identification method.
Barradas et al.~\cite{Barrados24_RAID} adapted the existing methodologies for C\&C traffic with TLS 1.3 protocol, which improves the TCP handshake protocol. They employed features related to the packet sizes, discarding all timing features since they are affected by the distance between client and server, as well as by network conditions.

Other works, have addressed the multi-classification problem to identify several kinds of attacks in IoT networks, \eg Denial of Service (DoS), and Port Scanning with several Machine Learning algorithms~\cite{Kumar2022_IoT,Churcher2021_Sensors} by considering both temporal and content-based features reaching accuracies in the order of 90\%.

Despite the importance of in-edge attack identification~\cite{Jia2020_IoT,Huong2021_Access}, and its efficient resource management, few existing works addresses the usage of green machine learning algorithms for network security applications~\cite{Ioannou2024_CC,Tekin2023_IoT}. The current proposed approaches (\eg TinyML\footnote{\url{https://github.com/mit-han-lab/tinyml}} techniques)  reduce the size of learning algorithms to be suitable for IoT devices, selecting one method over the other without a real optimization step based on the power usage.



\section{Methodology}\label{sec:methodology}

\begin{figure}[ht]
    \centering
    \includegraphics[width=1\linewidth]{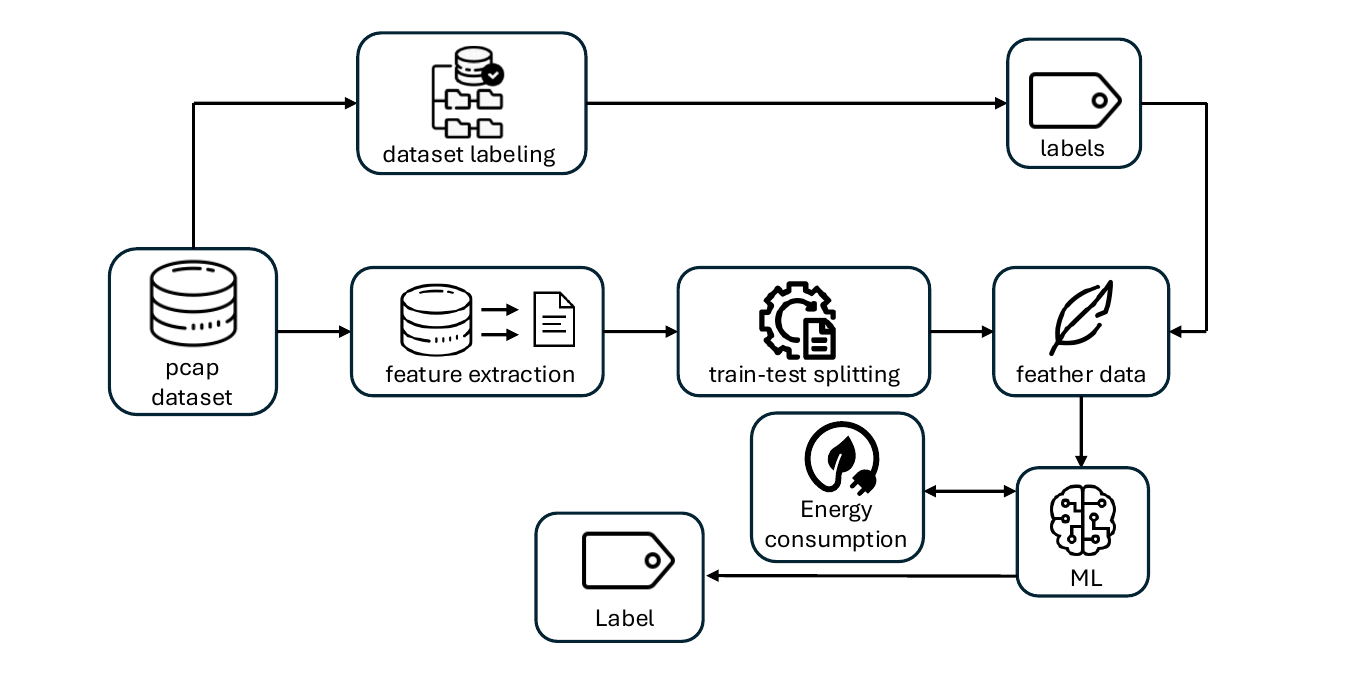}
    \caption{Training workflow with the basic blocks of the methodology: dataset labeling, feature extraction, training, and energy-based optimization.}
    \label{fig:workflow}
\end{figure}

This section introduces the proposed methodology to optimize ML models based on both energy consumption and performance, as depicted in Figure~\ref{fig:workflow}. Specifically, Section~\ref{sec:methodology:dataset} describes the employed dataset, Section~\ref{sec:methodology:features} discusses the extracted features, and Section~\ref{sec:methodology:modeltraining} introduces the model training approach with the optimization strategy. 

\subsection{Dataset}\label{sec:methodology:dataset}

To evaluate the approach, we used the Aposemat IoT-23 dataset~\cite{IoT23-dataset} by Garcia et al., a freely available labeled dataset of multiple {\tt PCAP} network capture files. The dataset contains benign and malicious traffic involving a variety of IoT sources (\eg smart hubs, smart lights, door lock devices) and malware. 

The \lstinline|NFStream|\footnote{\url{https://www.nfstream.org/}} tool was employed to analyze the PCAP files, obtain the label of the flows, and extract several meaningful statistical features, computed when a flow is closed, (\ie utilizing the totality of a connection), since our approach is tailored to \emph{post-mortem} traffic analysis. 
To identify each flow in the {\tt PCAP}, we employed the five-tuple \lstinline|(source IP address, source IP port, destination IP address, destination IP port, timestamp)| to match the identifiers of the original dataset in the file labels. The dataset was originally labeled by the authors Garcia et al. using the network traffic analyzer \lstinline|Zeek|\footnote{\url{https://zeek.org/}}.
Moreover, to remove noise, we dropped the flows with multiple contrasting labels (\eg a flow belonging both to a {\tt Mirai} and {\tt Kenjiro} botnet), or with no existing match in the \lstinline|Zeek| results.

Table~\ref{tab:dataset} shows the list of flow counts per class. Our processed dataset contains malicious traffic from seven malware tools: {\tt Kenjiro}, {\tt Mirai}, {\tt Hakai}, {\tt IRCBOT}, {\tt Hajime}, {\tt Hide\&Seek}, and {\tt Muhstick}, which represent the majority of malware found in real-world scenarios~\cite{Wang21_IoT}.
As noticeable, the majority of malware samples belong to the port-scan attack and the dataset is unbalanced in the malware/benign flows ratio. However, as it will be detailed in Section ~\ref{sec:results:error}, we reduced it by removing this class of attack, making the dataset nearly balanced and obtaining better results.



\begin{table}[ht]
    \centering
    \begin{tabular}{lp{0.2cm}lr}
    \toprule
    \textsc{\textbf{class}} & & \textsc{\textbf{family}} & \textsc{\textbf{flows}}\\
    \midrule
    \multirow{12}{*}{Malicious} & & IRCBOT -- port scan & 3627968\\
    & & Kenjiro -- port scan & 3525075\\
    & & Mirai -- port scan & 3236207\\
    & & Hajime -- port scan & 506947\\
    & & Hide and Seek -- port scan & 9558\\
    & & Muhstik -- port scan & 3671\\
    \cmidrule(lr){2-4}
    & & Mirai -- C\&C & 559\\
    & & Hakai -- C\&C & 103\\
    \cmidrule(lr){2-4}
    & & Kenjiro -- DoS & 776087\\
    & & Mirai -- DoS & 18344\\
    & & Muhstik -- DoS & 298\\
    \cmidrule(lr){1-4}
    Benign & & - & 1532194\\
    \bottomrule
    \end{tabular}
    \caption{Flow count grouped by class and attack types, \ie port scan, C\&C and DoS attacks.}
    \label{tab:dataset}
\end{table}

\subsection{Extracted Features}~\label{sec:methodology:features}

\lstinline|NFStream| was also employed to extract various statistical features, listed in Table~\ref{tab:features} for each correctly labeled flow. 
These features were chosen since using statistics computed per flow in combination with Machine Learning techniques to identify attacks is a time-proven approach in the scientific literature~\cite{Canavese22_CEE,Bader22_CCNC}.
Specifically, we computed each feature for three representations: \emph{i}) \textit{bidirectional}, which includes packets exchanged in both directions of the communication, \ie source and destination; \emph{ii}) \textit{source-to-destination}, where features are calculated solely on packets sent from the source to the destination; and \emph{iii}) \textit{destination-to-source}, which focuses on packets flowing in the reverse direction.
We purposely omitted the IP addresses and ports from the list of features for training and testing our models since these values are either meaningless or easy to spoof and especially to train a more generic model adaptable to every situation, while preserving privacy.

\begin{table}[ht]
    \centering
    \begin{tabular}{ll}
    \toprule
    \textsc{\textbf{feature}} & \textsc{\textbf{unit}}\\
    \midrule
    protocol\\
    IP version\\
    \cmidrule(lr){1-2}
    flow duration                                & \multirow{5}{*}{ms}\\
    maximum packet inter-arrival time            & \\
    minimum packet inter-arrival time            & \\
    mean packet inter-arrival time               & \\
    standard deviation packet inter-arrival time & \\
    \cmidrule(lr){1-2}
    transmitted bytes                            & \multirow{5}{*}{bytes}\\
    maximum packet size                          & \\
    minimum packet size                          & \\
    mean packet size                             & \\
    standard deviation packet size               & \\
    transmitted packets                          & packets\\
    \cmidrule(lr){1-2}
    TCP packets with ACK set                     & \multirow{8}{*}{packets}\\
    TCP packets with CWR set                     & \\
    TCP packets with ECE set                     & \\
    TCP packets with FIN set                     & \\
    TCP packets with PSH set                     & \\
    TCP packets with RST set                     & \\
    TCP packets with SYN set                     & \\
    TCP packets with URG set                     & \\
    \bottomrule
    \end{tabular}
    \caption{Statistical features employed in our methodology. Each statistic has been computed for the bidirectional, source-to-destination, and destination-to-source communications.}
    \label{tab:features}
\end{table}

Our statistics are based solely on packet timings, and they are computed only by analyzing the IP and TCP headers, which identify a flow in a communication. Moreover, the IP payload is not processed, and, therefore, this approach allows us to be encryption-agnostic with two main advantages. First, our methodology works equally well when the IP payload is encrypted (\eg with a TLS or DTLS connection). Second, our approach safeguards the users' privacy since the content of the IP packets is never inspected.

Once we built our dataset, we randomly split it into a training and a test set following an 80-20 ratio.

\subsection{Model training}\label{sec:methodology:modeltraining}

In our experiments, we tested three of the most employed Machine Learning algorithms for network security applications~\cite{Canavese22_CEE,Bader22_CCNC} using the well-known Python package \lstinline|scikit-learn|\footnote{\url{https://scikit-learn.org/}}. In particular, the models are tree-based techniques: \emph{i}) a \emph{decision tree}, or single-tree~\cite{Navada2011_CSG}, which is a hierarchical model that recursively splits data based on feature conditions to make predictions. Internal nodes represent decisions, and leaf nodes represent outcomes; a \emph{ii}) \emph{random forest}~\cite{Liu2012_ICA} that is an ensemble of multiple decision trees that usually shows improved accuracy over a single tree by carefully deciding how to split the nodes; and \emph{iii}) \emph{extra-trees} (Extremely Randomized Trees)~\cite{Geurts2006_ML}, which are also ensembles of trees, but they split the nodes randomly.

For each algorithm, we leveraged \lstinline|optuna|\footnote{\url{https://optuna.org/}}, a well-known hyperparameter optimization framework that helps to automate parameter search, to train four versions of each model given a function to optimize: 

\begin{itemize}
    \item a \emph{default} model, that is, the model trained with the default hyperparameters of \lstinline|scikit-learn|;
    \item a \emph{max green} model, that is, the optimized model with the lowest energy consumption at testing time;
    \item a \emph{max MCC} model, that is, the optimized model with the highest MCC (Matthew's Correlation Coefficient\footnote{\url{https://scikit-learn.org/stable/modules/generated/sklearn.metrics.matthews_corrcoef.html}});
    \item a \emph{balanced} model, that is, the model obtained with a multi-objective optimization to maximize the MCC and minimize the energy consumption. 
    Due to how \lstinline|optuna| works, we might encounter multiple optimal models. As specified later in Section~\ref{sec:results:optimization}, we selected the best model as the one that is geometrically closest to the point $(0, 1)$ in the Pareto front, where the first value is the power consumption and the second is the MCC. In other words, a model offering good discriminating capabilities and power saving without sacrificing too much of the two metrics.
\end{itemize}

To compute consumed energy, we employed {\tt Carbon Tracker} which is, as outlined in Secion~\ref{sec:sota:energy_measurement}, the best tool at the state of the art able to estimate the actual power usage.

Model optimization was performed based on energy consumption and performance during the testing phase. That is because the ultimate goal is to lower resources for the running algorithm on-premises while the training is performed on the server to account for large datasets. Additionally, due to class imbalance, model performance is computed with the Matthew's Correlation Coefficient that considers all four values of the confusion matrix, \ie True Positives (TP), True Negatives (TN), False Positives (FP), and False Negatives (FN).
\begin{table}[h]
    \centering
    \begin{tabular}{ll}
         \toprule
         CPU & Intel\textregistered{} Core\texttrademark{} i9-11950H CPU @ 2.60GHz\\
         RAM & 32 GiB\\
         \cmidrule(lr){1-2}
         OS & Debian GNU/Linux\\
         kernel & 6.2.10\\
         Python & 3.13.1\\
         \lstinline|scikit-learn| & 1.6.1\\
         \lstinline|optuna| & 4.2.0\\
         \lstinline|CarbonTracker| & 2.0.1\\
         \bottomrule
    \end{tabular}
    \caption{Specifics of our experimental setup.}
    \label{tab:machine}
\end{table}

\section{Results}\label{sec:results}

This Section discusses the results we obtained. In particular, Section~\ref{sec:results:expsetup} outlines the experimental setup we employed, Section~\ref{sec:results:overview} gives an overview of the obtained general results, Section~\ref{sec:results:optimization} concerns the hyperparameter tuning results, and Section~\ref{sec:results:error} examines the False Negative samples.

\begin{table}[h]
    \begin{subtable}[t]{\textwidth}
        \centering
        \begin{tabular}{lp{0.2cm}lc|c|c}
            \toprule
            \textsc{\textbf{type}} & & \textsc{\textbf{version}} & \multicolumn{3}{c}{\textsc{\textbf{Hyperparameter}}}\\ \midrule
            \multirow{5}{*}[-1em]{single-tree} & & & \textsc{\textbf{max\_depth}} & \textsc{\textbf{min\_leaf}} & \textsc{\textbf{min\_split}}\\ \midrule
            & & default & $\infty$ & 1 & 2\\
            & & max green & 1 & 3 & 9\\
            & & max MCC & 14 & 5 & 29\\
            & & balanced & 13 & 5 & 13\\
            \bottomrule
        \end{tabular}
        \caption{Hyperparameters for single-tree classifiers.}\label{tab:hypeparam:t}
    \end{subtable}
    
    \begin{subtable}[t]{\textwidth}
        \centering
        \begin{tabular}{lp{0.2cm}lc|c|c|c|c}
            \toprule
            \textsc{\textbf{type}} & & \textsc{\textbf{version}} & \multicolumn{5}{c}{\textsc{\textbf{Hyperparameter}}}\\ \midrule
            \multirow{5}{*}[-1em]{random forest} & & & \textsc{\textbf{max\_depth}} & \textsc{\textbf{min\_leaf}} & \textsc{\textbf{min\_split}} & \textsc{\textbf{max\_feat}} & \textsc{\textbf{estim.}}\\ \midrule
            & & default & $\infty$ & 1 & 2 & \lstinline|sqrt| & 100\\
            & & max green & 71 & 7 & 29 & 14 & 10\\
            & & max MCC & 11 & 6 & 17 & 11 & 133\\
            & & balanced & 17 & 6 & 20 & 7 & 18\\
            \bottomrule
        \end{tabular}
        \caption{Hyperparameters for random forest classifiers.}\label{tab:hypeparam:rf}
    \end{subtable}

    \begin{subtable}[t]{\textwidth}
        \centering
        \begin{tabular}{lp{0.2cm}lc|c|c|c|c}
            \toprule
            \textsc{\textbf{type}} & & \textsc{\textbf{version}} & \multicolumn{5}{c}{\textsc{\textbf{Hyperparameter}}}\\ \midrule
            \multirow{5}{*}[-1em]{extra trees} & & & \textsc{\textbf{max\_depth}} & \textsc{\textbf{min\_leaf}} & \textsc{\textbf{min\_split}} & \textsc{\textbf{max\_feat}} & \textsc{\textbf{estim.}}\\ \midrule
            & & default & $\infty$ & 1 & 2 & \lstinline|sqrt| & 100\\
            & & max green & 169 & 6 & 18 & 6 & 10\\
            & & max MCC & 18 & 2 & 20 & 23 & 205\\
            & & balanced & 14 & 2 & 18 & 24 & 204\\
            \bottomrule
        \end{tabular}
        \caption{Hyperparameters for extra-trees classifiers.}\label{tab:hypeparam:et}
    \end{subtable}
    \caption{Hyperparameter chosen by \lstinline|optuna| for each model.}\label{tab:hypeparam}
\end{table}


\subsection{Experimental setup}\label{sec:results:expsetup}


The experiments were conducted on a machine equipped with the specifications in Table~\ref{tab:machine}. In this preliminary work, we employed a server machine for both the training and testing phases. First, the server is used in the training for large-scale datasets, which need many resources in terms of memory (RAM), CPU/GPU usage, storage capabilities, and battery life that IoT devices do not support. We preferred to use the server even in the testing phase to ensure reproducibility, precise power measurement, and full control of the experimental setup. As there is a wide variety of different IoT devices and some of them have proprietary systems, achieving large reproducibility, adaptability, and comparison of the results on different systems is difficult because of the different available environments.

In fact, as mentioned in Section~\ref{sec:results:overview}, the consumed energy remains low (in the order of $\mu$Wh), which is in line with the constrained resources of IoT devices~\cite{Atzori2010_CN}. Therefore, while the experiments were conducted on a server, the methodology remains compatible with resource-constrained environments.



\begin{table}[h]
    \centering
    \begin{tabular}{lp{0.2cm}lp{0.2cm}rrrr}
        \toprule
        \textsc{\textbf{type}} & & \textsc{\textbf{version}} & & \textsc{\textbf{$\mu$Wh}} & \textsc{\textbf{MCC}} & \textsc{\textbf{b. acc.}} & \textsc{\textbf{F1}}\\
        \midrule
        \multirow{4}{*}{single-tree} & & default &  & 19.35 & 0.52 & 94.70 & 75.46\\
        & & max green &  & 6.50 & 0.23 & 89.05 &  6.42\\
        & & max MCC   &  & 8.13 & 0.60 & 95.35 & 76.11\\
        & & balanced  &  & 7.93 & 0.60 & 95.32 & 72.10\\
        \cmidrule(lr){1-8}
        \multirow{4}{*}{random forest} & & default &  & 299.83 & 0.52 & 94.68 & 75.76\\
        & & max green &  & 22.72 & 0.58 & 95.24 & 76.70\\
        & & max MCC   &  & 124.90 & 0.61 & 95.33 & 73.85\\
        & & balanced  &  & 28.70 & 0.61 & 95.33 & 74.20\\
        \cmidrule(lr){1-8}
        \multirow{4}{*}{extra-trees} & & default &  & 284.45 & 0.526 & 94.69 & 75.64\\
        & &  max green &  & 22.04 & 0.57 & 95.09 & 72.26\\
        & & max MCC   &  & 213.38 & 0.61 & 95.33 & 74.38\\
        & & balanced  &  & 49.68 & 0.60 & 95.24 & 74.08\\
        \bottomrule
    \end{tabular}
    \caption{Performance statistics of our classifiers. In particular, for each model, performance is shown in terms of average $\mu$Wh per test sample, Matthews' Correlation Coefficient (MCC), Balanced Accuracy, and F1-score.}
    \label{tab:results:before}
\end{table}

\subsection{Overview}\label{sec:results:overview}

Tables~\ref{tab:hypeparam} and~\ref{tab:results:before} report the selected hyperparameters by \lstinline|optuna| for each of the trained models, and the average energy consumption per testing sample and model performance in terms of Matthews' Correlation Coefficient (MCC), balanced accuracy, and F1-score to account for class imbalance in the dataset. These scores are chosen to consider both the dataset imbalance and the need to keep false negatives low to avoid malicious flows going undetected.

Interestingly, the default models offer strong discriminating power, reaching about 99\% balanced accuracy and 76\% F1-score. However, they are always the most energy-hungry, consuming about 13 times more $\mu$Wh than their max green counterparts. This is consistent with the default hyperparameters (see Table~\ref{tab:hypeparam}) in \lstinline|scikit-learn|, which were chosen to offer good performance in multiple scenarios, completely ignoring power consumption. 

As expected, the max green models are the most eco-friendly. However, this version of the single-tree has poor performance with 6\% of F1-score and 0.239 of MCC. This suggests that the green model has a weak correlation between its prediction and the label class. 
These relatively poor results may be caused by \lstinline|optuna| selecting a max depth of 1 (see Table~\ref{tab:hypeparam:t}), meaning that the model makes a single split based on one feature, oversimplifying the classification. 
Instead, the other versions of the single-tree seem to be the most efficient and high-performing, even with respect to the other models, \ie random-forest and extra-trees, which still offer good performance statistics but are highly resource-demanding due to the ensemble nature of these classifiers, which aggregate multiple weak learners, requiring more CPU and memory.

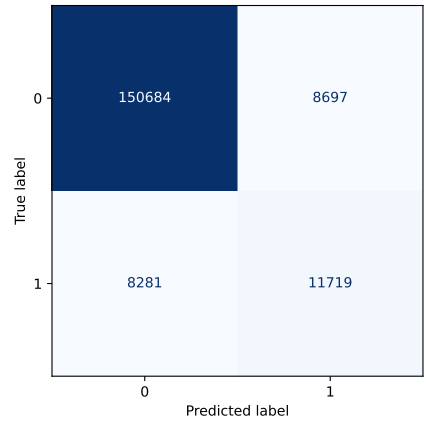
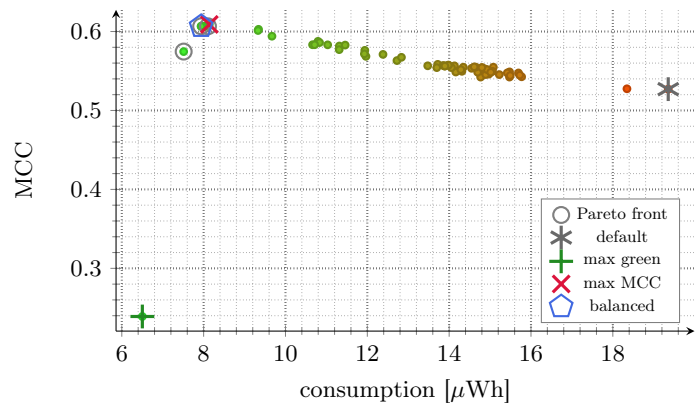
\begin{figure}[htbp]
    \centering
    \begin{subfigure}[h]{0.7\textwidth}
        \includegraphics[width=\linewidth]{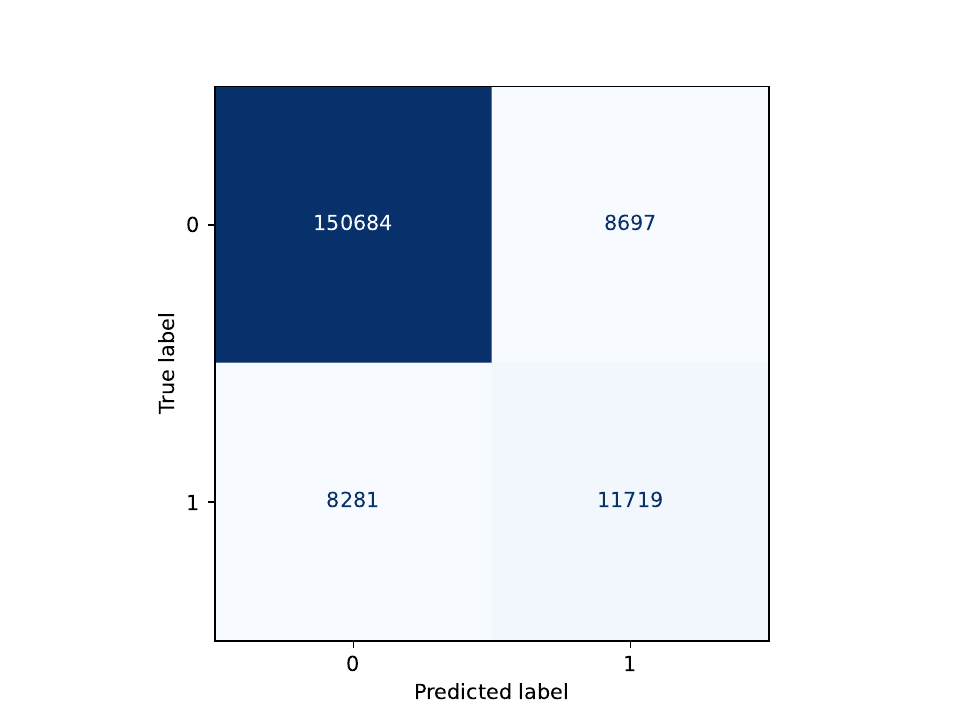}
        \caption{Simplified Confusion Matrix of the balanced single-tree model.}
        \label{fig:CM_balanced_tree}
    \end{subfigure}
    
    \begin{subfigure}[h]{0.8\textwidth}
    \centering
    \pgfplotsset{colormap={GreenToRed}{rgb255(0cm)=(50, 255, 50); rgb255(1cm)=(255, 69, 0)}}
    \begin{tikzpicture}
        \begin{axis}[axis lines=left, enlargelimits=0.05, grid=both, minor tick num=4, grid style={DimGray, very thin, densely dotted}, major grid style={Black, thick}, minor grid style={}, every axis plot/.append style={thick}, width=0.95\linewidth, height=0.6\linewidth, legend style={at={(0.975, 0.025)}, draw=Gray, anchor=south east, font=\footnotesize, nodes={scale=0.7, transform shape}}, xlabel={consumption [$\mu$Wh]}, ylabel=MCC]
        \addplot[scatter, only marks, mark=*, mark size=1.25pt, colormap name=GreenToRed, point meta=x] table {
            12.383448309143539 0.5710188966046248
            15.080129597640603 0.5548171283524592
            15.770920237017753 0.5426602961111936
            15.05979415385996 0.5495838095876698
            7.51204200136588 0.5745833126450248
            15.69339668330618 0.5475159254947736
            8.010060585476069 0.6065738205382468
            11.93888304971154 0.5761382774400713
            11.316440084499892 0.5816397497243361
            15.476103218173307 0.5488616326095312
            11.458205935997826 0.5828083702540539
            14.803534699245166 0.554262404941609
            14.345620341842988 0.5548134269732943
            13.721208762171841 0.5584135072301788
            10.794304567201774 0.587542398726148
            13.853348279575036 0.5569940069968293
            14.129834936541842 0.5565168893522353
            14.867013332881909 0.5527777786225528
            18.345178876294263 0.5274531120759743
            14.801912971905871 0.5452248408719601
            13.47667945807244 0.556577101800072
            11.966944277328425 0.5685249111129208
            14.311730287935015 0.5495993992901934
            12.724442799906894 0.5631908705618918
            14.162186867395633 0.5486966721937314
            13.697775351677528 0.5542569402579358
            14.769103138538823 0.5422676110883188
            8.105890267486604 0.6065935244168837
            11.031226277096383 0.5831297417820415
            14.697082549524392 0.5484304905927357
            11.927737346775029 0.5717019529887378
            14.96888245633752 0.5460229978581245
            9.669896809333064 0.5940539531169031
            14.948518311245861 0.5510300221403323
            10.662257594893395 0.5832418487143065
            14.927866267617304 0.5483681312572329
            15.226275297030389 0.5454402378119266
            14.916642946384888 0.5511781407727122
            14.547558546988835 0.5534039718474406
            11.312886963170062 0.5770462309576149
            14.901105724713183 0.5483930318037933
            14.893617753319287 0.5454637260425167
            13.888902316564286 0.5562526927827195
            14.056391759127964 0.5542821538593192
            9.341141168834765 0.6027814512484474
            14.648617831033212 0.5548714006155999
            15.724598708993714 0.5447529177154423
            13.98469858729322 0.5574084790728931
            15.473897229306534 0.5422867115208246
            14.929149800302234 0.5517130140417111
            15.015901262440003 0.5483930816299769
            13.978378760264224 0.5573744177115898
            8.028367354011152 0.6064714098917363
            9.329926287189606 0.601281223623593
            14.611839903454008 0.5552115805923462
            12.834052294141612 0.5673193328563113
            15.421651587307142 0.5476047479781407
            14.285070221539069 0.5517047367898632
            13.896106863477858 0.5562519543027696
            10.832697651564926 0.586006613884263
            15.492231180934715 0.5441847673749294
            10.712413025887725 0.5832548256731139
            7.9394830350169405 0.6066392938884424
            14.799214032890545 0.5542629866030386
            19.354 0.527
            6.502 0.239
            8.139 0.609
            7.9394830350169405 0.6066392938884424
        };
        \addplot[only marks, mark=o, mark size=3pt, thick, draw=Gray] table {
            7.51204200136588 0.5745833126450248
            7.9394830350169405 0.6066392938884424
            8.105890267486604 0.6065935244168837
        };
        \addplot[only marks, mark=asterisk, every mark/.append style={very thick}, mark size=4.5pt, thick, draw=DimGray] table {
            19.354 0.527
        };
        \addplot[only marks, mark=+, every mark/.append style={very thick}, mark size=4.5pt, thick, draw=ForestGreen] table {
            6.502 0.239
        };
        \addplot[only marks, mark=x, every mark/.append style={very thick}, mark size=4.5pt, thick, draw=Crimson] table {
            8.139 0.609
        };
        \addplot[only marks, mark=pentagon, mark size=4.5pt, thick, draw=RoyalBlue] table {
            7.9394830350169405 0.6066392938884424
        };
        \legend{, Pareto front, default, max green, max MCC, balanced}
        \end{axis}
    \end{tikzpicture}
        \caption{Pareto front for the single-tree models. The x-axis shows the mean energy consumption in terms of $\mu$Wh, and the y-axis is the MCC score.}
        \label{fig:pareto_tree}
    \end{subfigure}
    \caption{Confusion Matrix and Pareto front for the single-tree models showing classification accuracy and the performance in relation to the consumed energy.}
    \label{fig:tree_performance}
\end{figure}

\subsection{Optimization of the balanced models}\label{sec:results:optimization}

This section discusses the hyperparameter tuning process of the balanced version of the selected models. For each optimization, we asked \lstinline|optuna| to perform $64$ iterations, \ie the best number of iterations to achieve good optimization. The optimizations aimed to maximize the MCC and minimize the energy footprint during the inference phase. In particular, we refer only to the single-tree model, which was found to be the most efficient and to have the highest performance as outlined in Section~\ref{sec:results:overview}. Similar considerations can be made for the other models.

The Pareto front in Figure~\ref{fig:pareto_tree} helps in understanding how \lstinline|optuna| selects the best models. It shows all the single-tree models tested by \lstinline|optuna| in a scatter plot, where the points represent the Pareto front, the x-axis is the mean energy consumption per sample, and the y-axis represents the performance in terms of MCC. A solution is considered Pareto-optimal if no other configuration performs better in both objectives simultaneously~\cite{Tušar2015_TEC}. Points on the front are non-dominated, meaning improving one metric would negatively impact the other.

For example, increasing the MCC beyond a certain point may require a model that consumes significantly more energy, while reducing energy usage might come at the cost of lower classification accuracy. This provides a valuable decision boundary, allowing model selection based on application-specific priorities, such as maximizing accuracy, minimizing energy usage, or achieving a balanced compromise.

The default model (black star) presents high energy consumption ($\sim$19 $\mu$Wh) with moderate MCC ($\sim$0.53) and, as noticeable even from Table~\ref{tab:results:before}, it is not efficient but is a reference point for suboptimal optimizations of the model. 
The max green (green cross) has the lowest consumption ($\sim$7 $\mu$Wh), but low MCC ($\sim$0.4), \ie it is optimized only for energy consumption, suitable only if energy minimization is the priority, and accuracy is less critical. Conversely, the max MCC (red cross) is appropriate when maximizing the performance is critical, even though the consumed energy is not as low as for the green model ($\sim$8 µWh).
Finally, the balanced model (blue pentagon) is similar to the max MCC version with similar consumed energy and performance. Indeed, they both remain eco-friendly, maintaining the generalization capability. This means that both max MCC and balanced variants are suitable for running on an IoT device.

\begin{figure}[h]
    \centering
    \includegraphics[width=\linewidth]{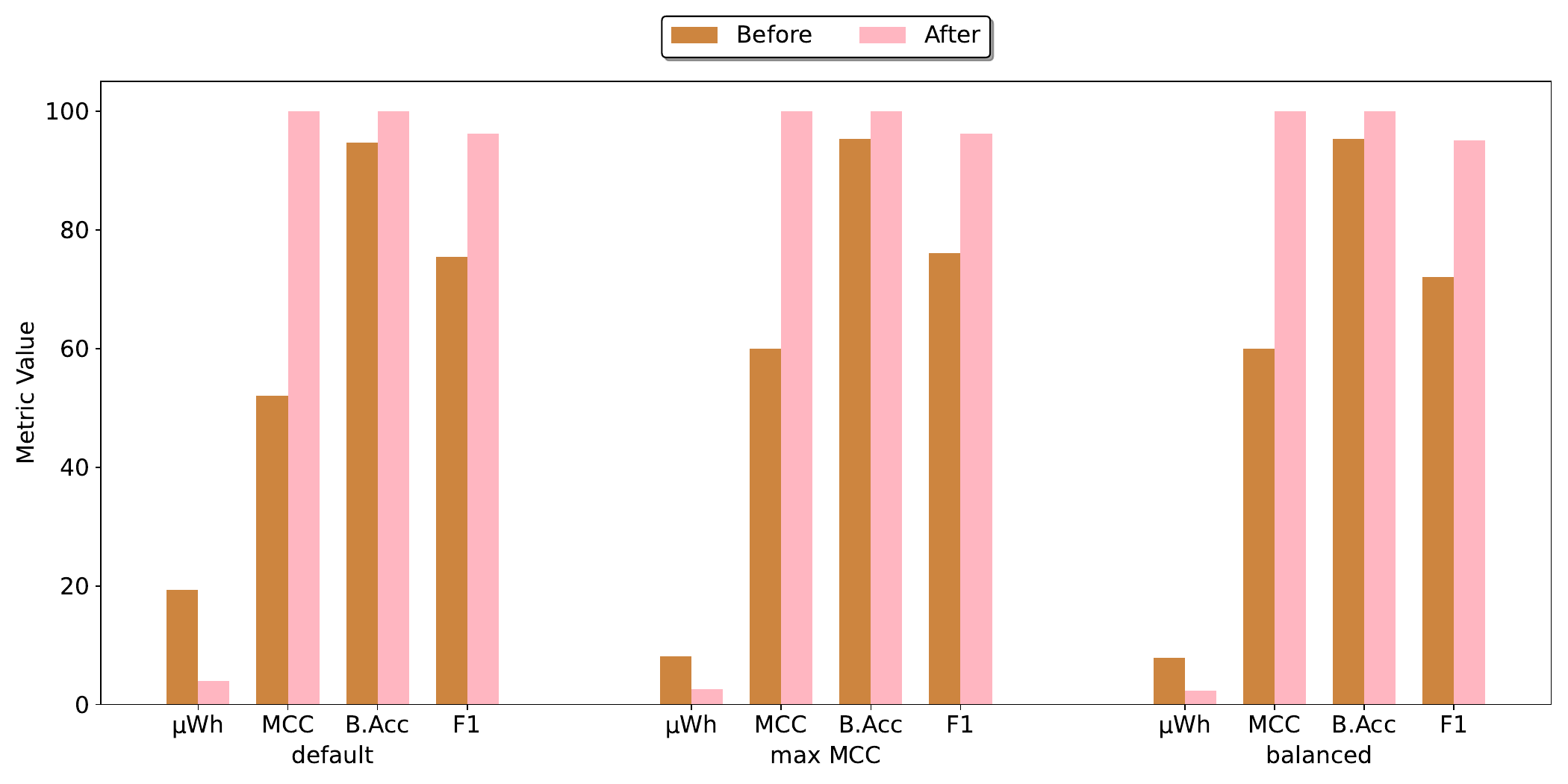}
    \caption{Comparison between performance of the optimized single-tree models (\ie default, max MCC, and balanced) before (left bars) and after (right bars) port-scanning attack removal. MCC is multiplied by 100 to be suitable for the graph.}
    \label{fig:comparison}
\end{figure}

\subsection{Error analysis}\label{sec:results:error}



\begin{figure}[h]
    \centering
    \pgfplotsset{colormap={GreenToRed}{rgb255(0cm)=(50, 255, 50); rgb255(1cm)=(255, 69, 0)}}
    \begin{tikzpicture}
        \begin{axis}[axis lines=left, enlargelimits=0.05, grid=both, minor tick num=4, grid style={DimGray, very thin, densely dotted}, major grid style={Black, thick}, minor grid style={}, every axis plot/.append style={thick}, width=0.95\linewidth, height=0.6\linewidth, legend style={at={(0.975, 0.025)}, draw=Gray, anchor=south east}, xlabel={consumption [$\mu$Wh]}, ylabel=MCC]
        \addplot[scatter, only marks, mark=*, mark size=1pt, colormap name=GreenToRed, point meta=x] table {
            2.39039921544828 0.9990339693585234
            2.430072328872272 0.8986993476834283
            2.3814788504792994 0.999068009576898
            2.364851380972048 0.9989284445795672
            4.154389190536782 0.9989267297643828
            2.4014697530726683 0.9738004205076097
            2.3952542194099307 0.9990983538574082
            2.3097872857859305 0.9987106370958225
            2.3440877613956843 0.998534207409976
            2.3778542439615 0.9941720656457309
            2.363760361034923 0.9990234213447919
            2.360430818832445 0.9992446060538589
            3.359562711161123 0.9338008066450076
            2.340623149565954 0.998927220777838
            2.363776108321893 0.9989032353514936
            2.355157664124037 0.9995017385262959
            2.3259346479525185 0.9954012858956194
            2.289037590847202 0.9992674108395918
            3.4038844489873976 0.9987835469158876
            2.3599433265746677 0.9989262488543648
            2.4156536100008985 0.9988603974707173
            3.6411540894760486 0.9988119255999361
            2.4045556885579646 0.9989274674049959
            2.3695814419288657 0.9991142731564735
            3.7257059116663522 0.9994889880826815
            2.3853908739471246 0.999116233649613
            2.4084909222605497 0.9991142719354189
            2.3353330605216307 0.998648502047465
            2.3685238727343774 0.9991159865289099
            2.32656192393597 0.9993453483256934
            2.406512514056981 0.9991015392121454
            3.3715095226668508 0.9990366452576563
            2.409387665084897 0.9941331213665564
            2.294667470679738 0.9991137828279543
            3.385650538839614 0.9987434197350552
            2.3298081574195275 0.9992441174897028
            2.3204387877917445 0.9989024927212979
            3.4296136394422025 0.9990467192297336
            2.305531325708202 0.9993698370043527
            2.343345229621459 0.9992490158967692
            2.282778396163576 0.9990493965675111
            2.3329290212903384 0.9990001659043107
            2.279808908878754 0.9992505043593024
            2.2784038392130705 0.9990290830589207
            3.428011393990701 0.9992483099755641
            2.3374385601681844 0.9989267327517134
            2.3387054734657293 0.9990241564093402
            2.3563122207564393 0.999236301194584
            2.3984036767535475 0.9988121705945788
            2.289105329182523 0.982095671160869
            3.4290617528763576 0.9989301584980405
            2.3213106898780085 0.9987123501461247
            2.301214076349789 0.9989487573143014
            2.3806107500259825 0.9988599059727771
            2.3584953507250837 0.9989957817937448
            2.4067278947379735 0.999002861208598
            2.413950617572511 0.9989260010327258
            2.403555742623723 0.9993355309246919
            2.4286325671766513 0.9954465897492664
            2.447969075162963 0.9991162322463488
            2.4141960750822338 0.9990209745247636
            2.44033416832438 0.9991473519471452
            2.4530540724579684 0.9991130478975244
            
            2.355157664124037 0.9995017385262959
            2.289037590847202 0.9992674108395918
            2.305531325708202 0.9993698370043527
            2.279808908878754 0.9992505043593024
            2.2784038392130705 0.9990290830589207
        };
        \addplot[only marks, mark=o, mark size=3pt, very thick, draw=Gray] table {
            2.355157664124037 0.9995017385262959
            2.289037590847202 0.9992674108395918
            2.305531325708202 0.9993698370043527
            2.279808908878754 0.9992505043593024
            2.2784038392130705 0.9990290830589207
        };
        \addplot[only marks, mark=asterisk, every mark/.append style={very thick}, mark size=4.5pt, very thick, draw=DimGray] table {
            4.026786218931780 0.9996341791152958
        };
        \addplot[only marks, mark=x, every mark/.append style={very thick}, mark size=4.5pt, thick, draw=Crimson] table {
            2.547078469123159 0.9996366331293386
        };
        \addplot[only marks, mark=pentagon, mark size=4.5pt, very thick, draw=RoyalBlue] table {
            2.355157664124037 0.9995017385262959
        };
        \legend{, Pareto front, default, max MCC, balanced}
        \end{axis}
    \end{tikzpicture}
    \caption{Pareto front for the tree models after removing port-scanning flows.}
    \label{fig:pareto_tree_after}
\end{figure}
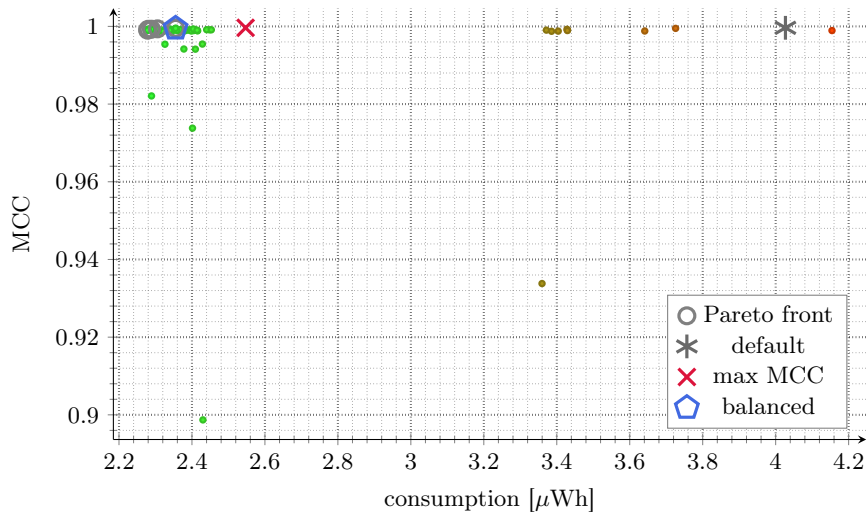

As outlined in Sections~\ref{sec:results:overview} and~\ref{sec:results:optimization}, single-tree model variants are the most efficient both in terms of performance and energy consumption. In particular, the balanced model, even though similar to the max MCC version, requires fewer resources and achieves comparable MCC. Figure~\ref{fig:CM_balanced_tree} shows the model confusion matrix, depicting that legitimate samples are well recognized, while it fails to clearly recognize a good portion of malicious flows—over $40$\% of actual true positives are misclassified. Therefore, even though MCC is good, \ie greater than 0, it is not good enough to be comparable with other works at the state-of-the-art~\cite{Davanian24_AsiaCCS} and requires attention due to the criticality of the application.

Analyzing the dataset and the corresponding features, we found that the models failed to recognize port-scan attacks whose flows are similar to legitimate ones. Therefore, we conducted a study of methodology performance while removing port-scan attacks from the dataset in Table~\ref{tab:dataset}, training only the single-tree model, which was the best performing of the three selected algorithms. We did not optimize the hyperparameters to achieve optimal energy consumption, as we expected it to have low accuracy.

Figure~\ref{fig:comparison} shows a comparison of single-tree model performance before and after the removal of port-scan attacks. As noticeable, MCC, Balanced Accuracy and F1 score increased after removing port-scan attacks while the average consumed $\mu$Wh reduced. 
The latter is expected because the less are the flows, the less complex the model is and therefore less resources are employed.

As before, we analyze the Pareto front in Figure~\ref{fig:pareto_tree_after}. It includes many points tightly clustered around low consumption ($\sim$$2.3$–$2.6$ $\mu$Wh) with near-perfect MCC, suggesting that it is possible to reduce energy without compromising the performance. For example, the max MCC (red cross) offers the absolute best MCC performance ($1.0$) but at a slightly higher energy cost, while the balanced model (blue pentagon) lies on the Pareto front, delivering strong performance ($0.995$) with modest energy use ($2.35$ $\mu$Wh). 
On the contrary, the default model (black star), despite achieving a high MCC ($0.9997$), is inefficient because it consumes significantly more energy than necessary, \ie $\sim50$\% more than the other two versions, for comparable accuracy, as discussed in Sections~\ref{sec:results:overview} and~\ref{sec:results:optimization}. 

\section{Conclusion}\label{sec:conclusion}

In this work, we tested the energy efficiency of tree-based Machine Learning algorithms trained to detect malicious network traffic generated by common IoT malware. The methodology is based on flow statistical features to preserve the privacy of legitimate communications. Additionally, we developed an optimization strategy with \lstinline|optuna| and \lstinline|Carbon Tracker| based on testing phase performance, considering both power consumption in Wh and MCC, while the training stage does not take power limitations into consideration. The reason is twofold. First, the trained detection algorithms can run on-premises, ideally on constrained IoT devices. Second, reducing the energy impact of the testing stage counteracts energy waste, increases the device operability lifecycle, and reduces the carbon footprint. 

We tested the methodology on a dataset consisting of three IoT attacks, reaching interesting results. Indeed, the models maintain high performance while keeping low energy consumption. The balanced version of the models, \ie models trained to balance both MCC and $\mu$Wh, attained about $0.60$ MCC and a reduction of $60-90\%$ in consumed resources compared to the models trained with the default hyperparameters. These results suggest that ML-based IDS systems are suitable for running on on-premise devices. Additionally, we studied the model errors and found that the dataset has biases with respect to port-scanning attacks, which have similar features to legitimate traffic flows.

However, the proposed approach still has some limitations. It is tested on only one dataset, limiting generalizability with respect to different network topologies and attacks. Additionally, the experimental setup lacks a constrained device to run the trained model to compute energy efficiency. Indeed, the optimization of the hyperparameters with respect to the testing phase results is done on the same server where the models are trained. 

In the future, we plan to improve our methodology with live analysis to test energy efficiency and inspect incoming network streaming. 
Additionally, we will test the approach with Deep Learning algorithms, which are inherently more energy-demanding than the tree-based algorithms we selected.


%
%
%
\bibliographystyle{plain}
\bibliography{main}
\end{document}